\documentclass{article}
\usepackage{spconf,amsmath,amssymb,graphicx,amsfonts,makecell,hyperref}
\usepackage{xcolor}

\title{DIFFRENT: A DIFFUSION MODEL FOR RECORDING ENVIRONMENT TRANSFER OF SPEECH}
\name{Jaekwon Im \qquad Juhan Nam
\thanks{This work was supported by the National Research Foundation of Korea (NRF) grant funded by the Korea government (MSIT) (No. RS-2023-00222383).}
}
\address{Graduate School of Culture Technology, KAIST, Republic of Korea}
\begin{document}
\ninept
\maketitle
\begin{abstract}

Properly setting up recording conditions, including microphone type and placement, room acoustics, and ambient noise, is essential to obtaining the desired acoustic characteristics of speech. In this paper, we propose Diff-R-EN-T, a Diffusion model for Recording ENvironment Transfer which transforms the input speech to have the recording conditions of a reference speech while preserving the speech content. Our model comprises the content enhancer, the recording environment encoder, and the diffusion decoder which generates the target mel-spectrogram by utilizing both enhancer and encoder as input conditions. We evaluate DiffRENT in the speech enhancement and acoustic matching scenarios. The results show that DiffRENT generalizes well to unseen environments and new speakers. Also, the proposed model achieves superior performances in objective and subjective evaluation. Sound examples of our proposed model are available online\footnote{\url{https://jakeoneijk.github.io/diffrent-demo}}.
\end{abstract}
 
\section{Introduction}
\label{sec:intro}
The acoustic characteristics of speech are determined by various recording conditions, such as the type and position of microphones, room acoustics, and  ambient noises. 
The proper conditions depend on the usage of the recorded speech. For instance, voice-overs or audiobooks require a clean environment with high-quality microphones and non-reverberant space. On the other hand, automated dialog replacement (ADR) demands post-production to replicate the acoustic qualities that provide environment information of the dialog. 
 
While its significance is indisputable, producing speech audio within a targeted recording environment requires considerable professional knowledge and effort. To tackle this problem, we introduce ``recording environment transfer" which transforms input speech so that it has the recording condition of a reference speech.

Previous studies of recording environment transfer mainly concentrate on a single target recording environment. One of the examples is speech enhancement, which converts speech audio recorded in a noisy environment to a clean voice. Conventional speech enhancement models aim to mitigate noise or reverberation. Recently, a number of studies have tackled many different distortions simultaneously to achieve a realistic recording environment \cite{voicefixer, hifi2}
. Despite the success of these works, the model that deals with a single target environment has a limitation in usability. 

Another case of recording environment transfer task is to simulate the conditions of a desired acoustic environment. Previous work implemented this by estimating or generating room impulse response (RIR) of the target environment \cite{blinddetectir, finsdetectir, chen2020soundspaces, image2reverb}, matching audio effects such as equalization (EQ) \cite{equalizationmatchgermain2016} or reverb \cite{reverbmatch}, or converting audio from a source microphone to that from a target microphone \cite{mic2mic}. However, most of them focused on only one recording condition.

This work aims to address general recording environment transfer that takes into account microphone type and placement, room acoustics, and ambient noise all at once. This holistic environment transfer was previously attempted by the acoustic matching model \cite{acousticmatching}, which matches reverberation, EQ distortion, and noise to the target environment using a reference speech that contains the recording environment. While they show promising results, this model has several limitations. First, it tends to be over-fitted to environments in the training set. As a result, its performance in generating audio in unseen recording environments is degraded. Second, they mainly focus on the case when the target environment is reverberant and noisy. Therefore, the performance of the model in transforming a noisy and reverberant environment into a clean environment is relatively unexplored. Lastly, the model was not capable of transferring realistic noises.

In this paper, we propose a unified environment transfer model that can faithfully change the acoustic characteristics of speech to arbitrary recording conditions. We implement it with a diffusion model which have shown impressive performance in generating images \cite{ddpm, palette} and audio \cite{diffwave, diffsvc} and thus we call it Diff-R-EN-T, a Diffusion model \cite{firstdiffusion} for Recording ENvironment Transfer. DiffRENT consists of three modules: the recording environment encoder, the content enhancer, and the diffusion decoder. The recording environment encoder extracts the recording environment embedding from a reference speech with the target environment. The content enhancer filters out the recording conditions of the source speech while preserving the speech content. The diffusion decoder generates the transformed speech with the target environment given the outputs of the recording environment encoder and the content enhancer. We show that DiffRENT effectively transfer any recording environment, disentangling the recording environment and the speech content well. We validate it in the speech enhancement and acoustic matching scenarios and show that the model achieves superior performance in both objective and subjective evaluation.   


\section{DIFFRENT}
\label{sec:methodology}

\begin{figure}[!t]
\begin{minipage}[b]{1.0\linewidth}
  \centering
  \centerline{\includegraphics[width=8.5cm]{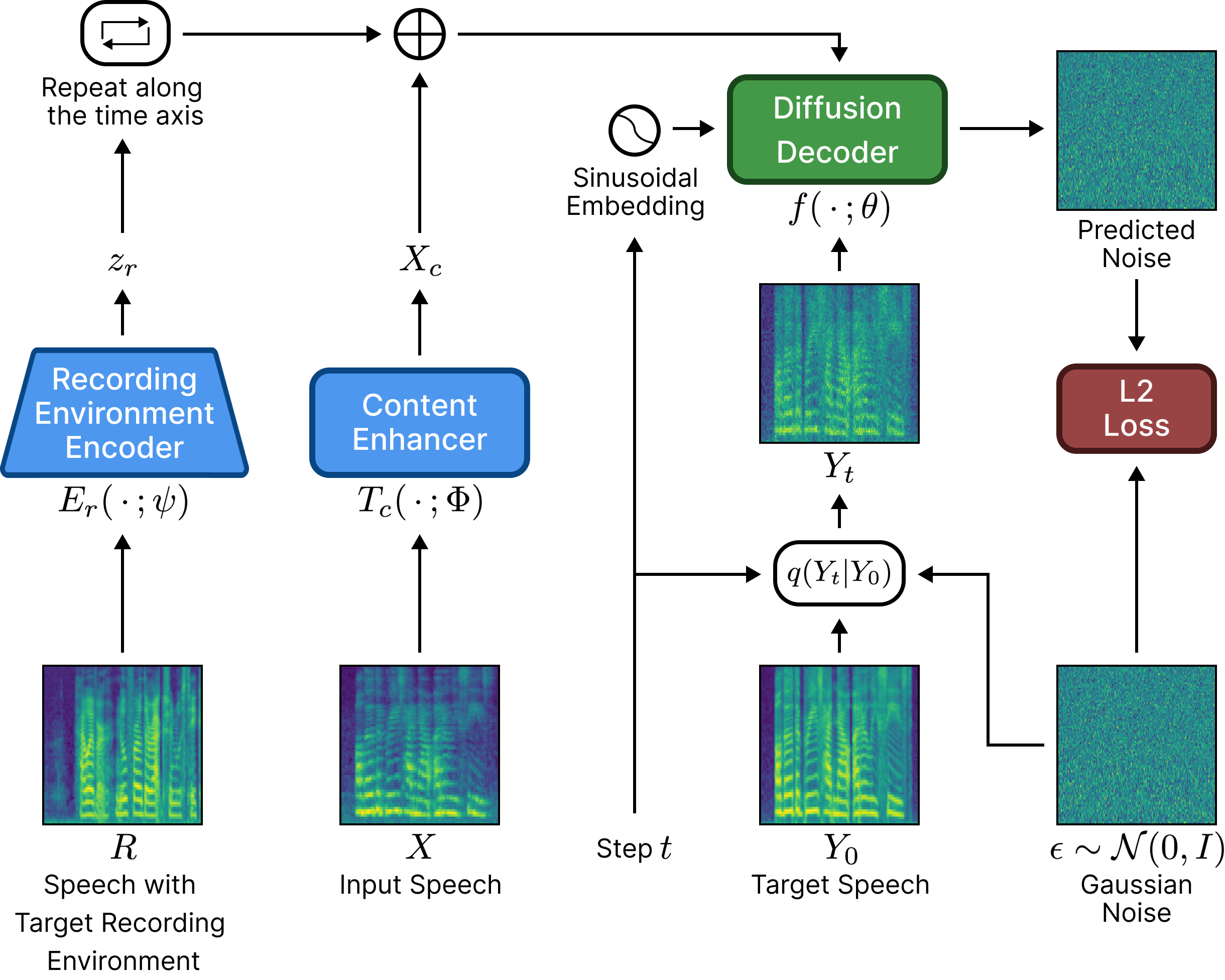}}
\end{minipage}
\caption{ The overall architecture of DiffRENT.}
\label{fig:overallarchitecture}
\end{figure}

Figure \ref{fig:overallarchitecture} illustrates the overview of the proposed model. Let $x \in \mathbb{R}^{L_1}$, $r \in \mathbb{R}^{L_2}$, and $y \in \mathbb{R}^{L_1}$ be content speech (input), reference speech in the target recording environment, and output speech with the content of $x$ and the recording environment of $r$, respectively, where $L_1$ and $L_2$ are the number of samples. 
Their log mel-spectrograms are represented as $X \in \mathbb{R}^{F \times T_1}$, $R \in \mathbb{R}^{F \times T_2}$, and $Y \in \mathbb{R}^{F \times T_1}$ where $F$ indicates the number of frequency bins, and $T_1$ and $T_2$ indicate the number of time frames. DiffRENT generates $Y$, given $X$ and $R$ as input conditions. The target audio signal $y$ can be obtained from $Y$ using a pre-trained HiFi-GAN vocoder \cite{hifigan}. The proposed model comprises the content enhancer $T_c( \, \cdot \, ;\Phi)$, the recording environment encoder $E_r(\, \cdot \,;\psi)$, and the diffusion decoder $f(\, \cdot \,;\theta)$. In the following, we describe each component in detail.

\subsection{Recording Environment Encoder}
The role of the recording environment encoder $E_r(\, \cdot \,;\psi)$ is to extract environmental acoustic features other than the speech content from the reference audio. We assume that the recording environment remains unchanged over time. From input mel-spectrograms $R$, $E_r(\, \cdot \,;\psi)$ produces embedding $z_r \in \mathbb{R}^C$, the disentangled recording environment representation. The diffusion decoder $f(\, \cdot \,;\theta)$ takes $z_r$ as an input condition. Without any other objectives, $E_r(\, \cdot \,;\psi)$ and $f(\, \cdot \,;\theta)$ are trained jointly with a denoising objective. We employ the ECAPA-TDNN \cite{ecapatdnn} architecture for the recording environment encoder. ECAPA-TDNN was originally designed to extract speaker embedding for speaker verification. Given its strong capability in capturing static features, we deduce its suitability for our task. We removed the AAM-softmax in ECAPA-TDNN and modified the number of nodes in the final fully-connected layer to align with the condition channel size of the diffusion decoder.

\subsection{Content Enhancer}
In order to preserve the speech content from the input audio and filter out the recording environment features, the content enhancer $T_c( \, \cdot \, ;\Phi)$ is trained by minimizing the following mean absolute error (MAE) loss:
\begin{equation}
    \mathcal{L} = \| X_{clean} - T_c(X;\Phi)\|_1
\end{equation}
where $X_{clean}$ is the log mel-spectrogram of clean speech having the same content as $x$. Note that the goal of $T_c( \, \cdot \, ;\Phi)$ is to facilitate improved content capturing for the diffusion decoder rather than speech enhancement. 
There are some specific artifacts or distortion in $X_c \in \mathbb{R}^{F \times T_1}$ as it is trained with the simple L1 loss. However, they are addressed in the diffusion decoder and so their influence on the overall model performance remains negligible. We employed a ResUNet from \cite{decouplingresunet} and slightly modified it to operate on mel-spectrogram as in \cite{resunetmel}.

\subsection{Diffusion Decoder}
To generate a target mel-spectrogram $Y_0 \sim p(Y | X_c,z_r)$, from the two input conditions $X_c$ and $z_r$, 
we apply a denoising diffusion probabilistic model \cite{ddpm}. A diffusion probabilistic model belongs to the category of generative models that learn a model distribution $p_{\theta}(Y_0)$ which approximates the data distribution $q(Y_0)$. A diffusion model, with the total number of diffusion steps $T$, is a latent variable model having the form $p_{\theta}(Y_0) = \int p_{\theta}(Y_{0:T})dY_{1:T}$, where 
$Y_1$, ..., $Y_T$ are latent variables of the same dimensionality as $Y_0$. It comprises two processes as described below. 


\subsubsection{Diffusion/Forward Process}
The diffusion/forward process is a fixed Markov process that gradually adds small Gaussian noise to the data over $T$ iterations, with the aim of making the distribution of $Y_T$ a standard Gaussian:

\begin{equation}
    q(Y_{1:T}|Y_0) := \prod_{t=1}^{T} q(Y_t|Y_{t-1})
\end{equation}
with $q(Y_t|Y_{t-1}) := \mathcal{N}(Y_t;\sqrt{1 - \beta_t}Y_{t-1},\beta_{t}I)$, where $\beta_t$ is a deterministic noise schedule constant satisfying $ t_1<t_2 \Rightarrow \beta_{t_1}<\beta_{t_2}$. The forward process at each step can be marginalized as follows:
\begin{equation}
    q(Y_t|Y_0) = \mathcal{N}(Y_t;\sqrt{\bar{\alpha}_t}Y_0, (1-\bar{\alpha}_t)I)
\end{equation}
where $\alpha_t := 1 - \beta_t$ and $\bar{\alpha}_t := \prod_{s=1}^{t}\alpha_s$.

\subsubsection{Reverse Process}
In DiffRENT, the reverse process 
can be defined as a Markov chain with learned Gaussian conditional transition distributions starting from $p(Y_T) = \mathcal{N}(Y_T;0,I)$.
\begin{equation}
    p_{\theta}(Y_{0:T}|X_c,z_r) := p(Y_T) \prod_{t=1}^{T} p_{\theta}(Y_{t-1}|Y_t,X_c,z_r)
\end{equation}
\begin{equation}
    p_{\theta}(Y_{t-1}|Y_t,X_c,z_r) := \mathcal{N}(Y_{t-1};\mu_{\theta}(Y_t,t,X_c,z_r),\tilde{\beta}_tI)
\end{equation}
where $\tilde{\beta}_t = \frac{1-\overline{\alpha}_{t-1}}{1-\overline{\alpha}_t}\beta_t$. The diffusion decoder is trained to optimize learnable parameters $\theta$, enabling the reverse of the diffusion process.

\subsubsection{Optimizing Diffusion Decoder}
To maximize the intractable likelihood $p_{\theta}(Y_0)$, evidence lower bound (ELBO) is used to train the diffusion decoder \cite{ddpm}. Given $Y$, $X$, and $R$, sampled from the training dataset, the objective function for training the diffusion decoder $f(\, \cdot \,, \theta)$ is as follows:
\begin{equation}
    \mathbb{E}_{(Y,X,R),t,\epsilon}  \lVert \epsilon - f(Y_t,t,T_c(X;\Phi),E_r(R;\psi);\theta) \rVert_{2}^{2}
\end{equation}
where $\epsilon \sim \mathcal{N}(0,I)$ and $t \in \{1, \cdots, T\}$ is sampled uniformly. The trainable parameters $\theta$ and $\psi$ are concurrently optimized while $\Phi$ is fixed.
\section{Experiments}
\label{sec:experiments}

\subsection{Dataset and Preprocessing}
We evaluated our proposed model on the DDS dataset \cite{ddsdataset} which includes 12 hours of clean speech data with 48 speakers (24 female and 24 male) and corresponding paired data re-recorded with different combinations of 3 microphones, 9 spaces, and 6 microphone positions. 
For the test set, we selected four speakers (two males and two females) and one recording environment setup (Uber Microphone, livingroom1, and F position). 

To enhance the robustness of content capturing, we implemented data augmentation by convolving clean speech with diverse impulse responses and adding noise. The noise samples were sourced from the REVERB Challenge database \cite{reverbchallenge}, and a total of 1207 impulse responses are obtained from the DetmoldSRIR dataset \cite{detmoldsrir} and the MIT Impulse Response Survey
Dataset \cite{mitdataset}. Note that data augmentation is exclusively applied to content speech, because the goal of the diffusion decoder is to learn the distribution of realistic environments rather than synthetic ones.

For our task, it is important that the neural vocoder operates in a speaker-independent manner and acquires the capability to synthesize speech in diverse recording environments. To achieve this, we trained the HiFi-GAN vocoder \cite{hifigan} with the DDS dataset \cite{ddsdataset} and  the LibriSpeech corpus \cite{librispeech}. The LibriSpeech corpus \cite{librispeech} initially designed for automatic speech recognition (ASR) research contains 982 hours of speech data from 2,484 speakers. Despite some considerable noisy audio files within this dataset, they align with the requirements of our task, as the vocoder is also required to generate noise effectively.

All audio sample rates 
were resampled to 16 kHz. Speech data were randomly segmented or padded to match the length of 4 seconds. The spectrogram was extracted from the audio by short-time Fourier transform (STFT) with a Hann window size of 1024 samples and a hop size of 256 samples. The log mel-spectrogram was computed through an 80-channel mel filterbank and the log magnitude compression. For the diffusion decoder, it was normalized to the range between -1 and 1 by a min-max normalization.

\subsection{Implementation Details}
\textbf{Diffusion Decoder:} We conducted experiments with two diffusion decoders: WaveNet \cite{wavenet} architecture in \cite{diffsvc} and U-Net architecture in \cite{palette}. The embedding $z_r$ channel size, contingent on the conditioning method of the diffusion decoder, was set to 256 for WaveNet and 80 for U-Net. We empirically found this difference in channel size does not yield a significant variance in the performance of the recording environment encoder. To align the time frame size of $z_r$ with $X_c$, $z_r$ is repeated along the time axis. In the case of U-Net, the final condition is established by concatenating $z_r$ with $X_c$. For the WaveNet architecture, the channel size of $X_c$ is transformed to 256 via a linear layer, followed by the summation with $z_r$ to make the input condition.

\noindent \textbf{Training Procedure:} 
We first trained the content enhancer for 225k steps. Then, the diffusion decoder and the recording environment encoder were trained jointly for 400k steps. We adopted a training and inference procedure similar to \cite{diffsvc} for the diffusion decoder. All models were trained with the AdamW optimizer and a batch size of 32. The learning rate started at 0.0008 and was halved every 20k steps.


\subsection{Evaluation}
We compare DiffRENT with the acoustic matching model \cite{acousticmatching}. Due to the absence of source code, we reproduced the model based on the description provided in \cite{acousticmatching}. To evaluate the speech enhancement capabilities of our model,  we also compare our method with CDiffuSE\cite{cdiffuse} and VoiceFixer \cite{voicefixer}, the state-of-the-art level speech enhancement model. The official implementations of them were employed. Comparison models were trained with the identical data configuration used in the proposed model.

We conducted an ablation study to evaluate the effectiveness of each component in DiffRENT. For the recording environment encoder, we used a simple encoder composed of 1D convolution with a kernel size of 1, followed by the attentive statistics pooling \cite{attentivepool} and batch normalization as a baseline model.
Furthermore, we evaluated the effect of the content enhancer with the models using the original input $X$ instead of the enhanced speech $X_c$.

\subsubsection{Objective Evaluation}
To assess the adaptability of our model in multiple tasks, we employed three test cases: 
\begin{enumerate}
   \item \textbf{Env-to-Clean:} evaluate the speech enhancement performance. 
  \item \textbf{Clean-to-Env:} evaluate the performance of transforming a clean environment into an unseen environment.
  \item \textbf{Env-to-Env:} evaluate the performance of transforming an unseen environment into a seen environment. 
\end{enumerate}
For each test case, 500 excerpt pairs composed of content speech $x$ and reference speech $r$ were chosen. For \textbf{Env-to-Env}, we randomly selected the target environment of each reference speech. We adopt the log-spectral distance (LSD) \cite{lsd} and structural similarity (SSIM) \cite{ssim} as the objective metrics. Furthermore, for \textbf{Env-to-Clean}, wideband perceptual evaluation of speech quality (PESQ-wb) \cite{pesq} and scale-invariant spectrogram to noise ratio (SiSPNR) \cite{voicefixer} are employed. SiSPNR is the scale-invariant signal-to-noise ratio (SiSNR) \cite{sisnr} computed on the magnitude spectrogram.

\subsubsection{Subjective Evaluation}
We conducted a subjective evaluation through a listening test with 20 participants for each test case. Each test case encompassed ten questions. For \textbf{Env-to-Clean}, participants were provided the content speech as a low anchor. They were instructed to evaluate two distinct criteria: content preservation and enhancement quality. For \textbf{Clean-to-Env} and \textbf{Env-to-Env}, participants were presented with the target speech as a high anchor, and they were asked to assess content preservation and environment similarity. In each question, participants assessed the audio files generated by our proposed model and the comparison model. Additionally, the target speech and the content speech were assessed to estimate upper and lower bounds. Note that content preservation evaluates how well speech content is preserved regardless of the recording environment. Therefore, both the target speech and the content speech (unprocessed) are expected to receive high scores in this criterion. We selected U-R2-C for all test cases. For \textbf{Env-to-Clean} and \textbf{Clean-to-Env} \& \textbf{Env-to-Env}, we selected VoiceFixer and the acoustic matching network as the comparison model, respectively. Each evaluation item was rated on a scale of 1 to 5 points.

\section{Results}

\begin{table}[t]
\centering
\footnotesize{
\begin{tabular}{lllll}
\textbf{Env-to-Clean}&        &   &      & \\ \Xhline{3\arrayrulewidth}
Method &   PESQ ↑    &SiSPNR ↑    &\multicolumn{1}{c}{LSD ↓}    &SSIM ↑  \\ \Xhline{3\arrayrulewidth}
Unprocessed &\multicolumn{1}{c}{1.34}&\multicolumn{1}{c}{8.45}   &\multicolumn{1}{c}{1.24} &\multicolumn{1}{c}{0.82}\\ 
Target-Mel&\multicolumn{1}{c}{2.9}&\multicolumn{1}{c}{13.83}&\multicolumn{1}{c}{0.23}&\multicolumn{1}{c}{0.99}\\ 
Target &\multicolumn{1}{c}{4.64}&\multicolumn{1}{c}{128.51}&\multicolumn{1}{c}{0.0}&\multicolumn{1}{c}{1.0}\\ \hline
A-Match \cite{acousticmatching}&\multicolumn{1}{c}{1.4}&\multicolumn{1}{c}{9.68}&\multicolumn{1}{c}{0.9}&\multicolumn{1}{c}{0.9}\\ 
CDiffuSE \cite{cdiffuse}&\multicolumn{1}{c}{1.32}&\multicolumn{1}{c}{9.51}&\multicolumn{1}{c}{0.94}&\multicolumn{1}{c}{0.87}\\ 
VoiceFixer \cite{voicefixer}&\multicolumn{1}{c}{1.6}&\multicolumn{1}{c}{\textbf{11.55}}&\multicolumn{1}{c}{\textbf{0.58}}&\multicolumn{1}{c}{\textbf{0.94}
}\\  \hline
W-R1   &\multicolumn{1}{c}{1.4}&\multicolumn{1}{c}{9.97}&\multicolumn{1}{c}{0.83}&\multicolumn{1}{c}{0.91}\\ 
W-R2 &\multicolumn{1}{c}{1.33}&\multicolumn{1}{c}{9.29}&\multicolumn{1}{c}{0.95}&\multicolumn{1}{c}{0.9}\\ 
W-R1-C   &\multicolumn{1}{c}{1.47}&\multicolumn{1}{c}{11.12}&\multicolumn{1}{c}{0.66}&\multicolumn{1}{c}{0.93}\\ 
W-R2-C &\multicolumn{1}{c}{1.53}&\multicolumn{1}{c}{11.24}&\multicolumn{1}{c}{0.64}&\multicolumn{1}{c}{0.93}\\ 
U-R1   &\multicolumn{1}{c}{1.63}&\multicolumn{1}{c}{10.72}&\multicolumn{1}{c}{0.75}&\multicolumn{1}{c}{0.92}\\ 
U-R2 &\multicolumn{1}{c}{\textbf{1.7}}&\multicolumn{1}{c}{10.92}&\multicolumn{1}{c}{0.69}&\multicolumn{1}{c}{0.92}\\ 
U-R1-C   &\multicolumn{1}{c}{1.47}&\multicolumn{1}{c}{11.12}&\multicolumn{1}{c}{0.66}&\multicolumn{1}{c}{0.93}\\ 
U-R2-C &\multicolumn{1}{c}{1.61}&\multicolumn{1}{c}{11.36}&\multicolumn{1}{c}{0.63}&\multicolumn{1}{c}{0.93}\\ 
\Xhline{3\arrayrulewidth}\\
\textbf{Clean-to-Env}     &   &   &\textbf{Env-to-Env}      &  \\ \Xhline{3\arrayrulewidth}
Method &\multicolumn{1}{c}{LSD ↓}   &\multicolumn{1}{c|}{SSIM ↑}   &\multicolumn{1}{c}{LSD ↓}  &SSIM ↑ \\ \Xhline{3\arrayrulewidth}
Unprocessed &\multicolumn{1}{c}{1.24}&\multicolumn{1}{c|}{0.82}&\multicolumn{1}{c}{0.73}&\multicolumn{1}{c}{0.87}\\ 
Target-Mel&\multicolumn{1}{c}{0.2}&\multicolumn{1}{c|}{0.98}&\multicolumn{1}{c}{0.21}&\multicolumn{1}{c}{0.98}\\ 
Target &\multicolumn{1}{c}{0.0}&\multicolumn{1}{c|}{1.0}&\multicolumn{1}{c}{0.0}&\multicolumn{1}{c}{1.0}\\  \hline
A-Match \cite{acousticmatching}&\multicolumn{1}{c}{0.82}&\multicolumn{1}{c|}{0.81}&\multicolumn{1}{c}{0.69}&\multicolumn{1}{c}{0.86}\\ \hline
W-R1   &\multicolumn{1}{c}{0.86}&\multicolumn{1}{c|}{0.84}&\multicolumn{1}{c}{0.68}&\multicolumn{1}{c}{0.87}\\ 
W-R2 &\multicolumn{1}{c}{1.0}&\multicolumn{1}{c|}{0.83}&\multicolumn{1}{c}{0.8}&\multicolumn{1}{c}{0.86}\\ 
W-R1-C   &\multicolumn{1}{c}{0.64}&\multicolumn{1}{c|}{0.85}&\multicolumn{1}{c}{0.62}&\multicolumn{1}{c}{0.87}\\ 
W-R2-C &\multicolumn{1}{c}{\textbf{0.58}}&\multicolumn{1}{c|}{\textbf{0.87}}&\multicolumn{1}{c}{0.57}&\multicolumn{1}{c}{0.89}\\ 
U-R1   &\multicolumn{1}{c}{0.76}&\multicolumn{1}{c|}{0.84}&\multicolumn{1}{c}{0.71}&\multicolumn{1}{c}{0.86}\\ 
U-R2 &\multicolumn{1}{c}{0.71}&\multicolumn{1}{c|}{0.86}&\multicolumn{1}{c}{0.59}&\multicolumn{1}{c}{0.89}\\ 
U-R1-C   &\multicolumn{1}{c}{0.64}&\multicolumn{1}{c|}{0.85}&\multicolumn{1}{c}{0.62}&\multicolumn{1}{c}{0.87}\\ 
U-R2-C &\multicolumn{1}{c}{0.59}&\multicolumn{1}{c|}{\textbf{0.87}}&\multicolumn{1}{c}{\textbf{0.55}}&\multicolumn{1}{c}{\textbf{0.9}}\\ 
\Xhline{3\arrayrulewidth}
\end{tabular}
}
\caption{Objective evaluation results. W: WaveNet decoder. U: U-Net decoder. R1: Baseline encoder. R2: Recording environment encoder based on ECAPA-TDNN. C: Content enhancer}
\label{table:objectice}
\end{table}

\subsection{Objective Evaluation}
Table \ref{table:objectice} shows the result of the objective evaluation. In all test cases, all proposed models except those using the WaveNet decoder without the content enhancer outperform the acoustic matching model in all metrics. In \textbf{Env-to-Clean}, three proposed models have better PESQ values than VoiceFixer, which performs best among the comparison models. Model U-R2-C achieves comparable performance compared to VoiceFixer across all metrics. In the majority of cases, the model employing the U-Net architecture in the diffusion decoder demonstrates superior performance compared to the model with the WaveNet decoder. Additionally, the models with R2 encoder consistently outperform the models with R1 encoder. It implies that enhanced disentanglement features contribute to the improved ability of the decoder to generate sound within the target environment. There is only a single case where the model with R1 outperforms the model with R2 (W-R1 and W-R2). This might be attributed to the limited capacity of the decoder, which could lead to degraded performance with diverse input conditions. The results also show the effectiveness of the content enhancer. 

\begin{table}
\centering
\footnotesize{
\begin{tabular}{lllllll}
 & \multicolumn{2}{l}{\textbf{Env-to-Clean}}  & \multicolumn{2}{l}{\textbf{Clean-to-Env}} & \multicolumn{2}{l}{\textbf{Env-to-Env}} \\ \Xhline{3\arrayrulewidth}
Method &\multicolumn{1}{c}{CP ↑} &  \multicolumn{1}{c|}{EQ ↑}&   CP ↑    &  \multicolumn{1}{c|}{ES ↑}&   CP ↑    &  ES ↑  \\ \Xhline{3\arrayrulewidth}
Unprocessed &\multicolumn{1}{c}{4.39}&\multicolumn{1}{c|}{2.31}&\multicolumn{1}{c}{4.35}&\multicolumn{1}{c|}{1.57}&\multicolumn{1}{c}{4.26}&\multicolumn{1}{c}{3.03} \\ 
Target &\multicolumn{1}{c}{4.72}&\multicolumn{1}{c|}{4.74}&\multicolumn{1}{c}{4.68}&\multicolumn{1}{c|}{4.66}&\multicolumn{1}{c}{4.71}&\multicolumn{1}{c}{4.60} \\ \hline
VoiceFixer \cite{voicefixer}&\multicolumn{1}{c}{3.52}&\multicolumn{1}{c|}{3.64}&\multicolumn{1}{c}{-}&\multicolumn{1}{c|}{-}&\multicolumn{1}{c}{-}&\multicolumn{1}{c}{-} \\
A-Match \cite{acousticmatching} &\multicolumn{1}{c}{-}&\multicolumn{1}{c|}{-}&\multicolumn{1}{c}{3.87}&\multicolumn{1}{c|}{2.92}&\multicolumn{1}{c}{4.02}&\multicolumn{1}{c}{3.45} \\
DiffRENT &\multicolumn{1}{c}{\textbf{4.22}}&\multicolumn{1}{c|}{\textbf{4.38}}&\multicolumn{1}{c}{\textbf{4.39}}&\multicolumn{1}{c|}{\textbf{4.00}}&\multicolumn{1}{c}{\textbf{4.33}}&\multicolumn{1}{c}{\textbf{4.15}} \\
\Xhline{3\arrayrulewidth}
\end{tabular}
}

\caption{Subjective evaluation results. CP: Content preservation. EQ: Enhancement quality. ES: Environment similarity}
\label{table:subjective}
\end{table}

\begin{figure}[!t]
\begin{minipage}[b]{1.0\linewidth}
  \centering
  \centerline{\includegraphics[width=8.3cm]{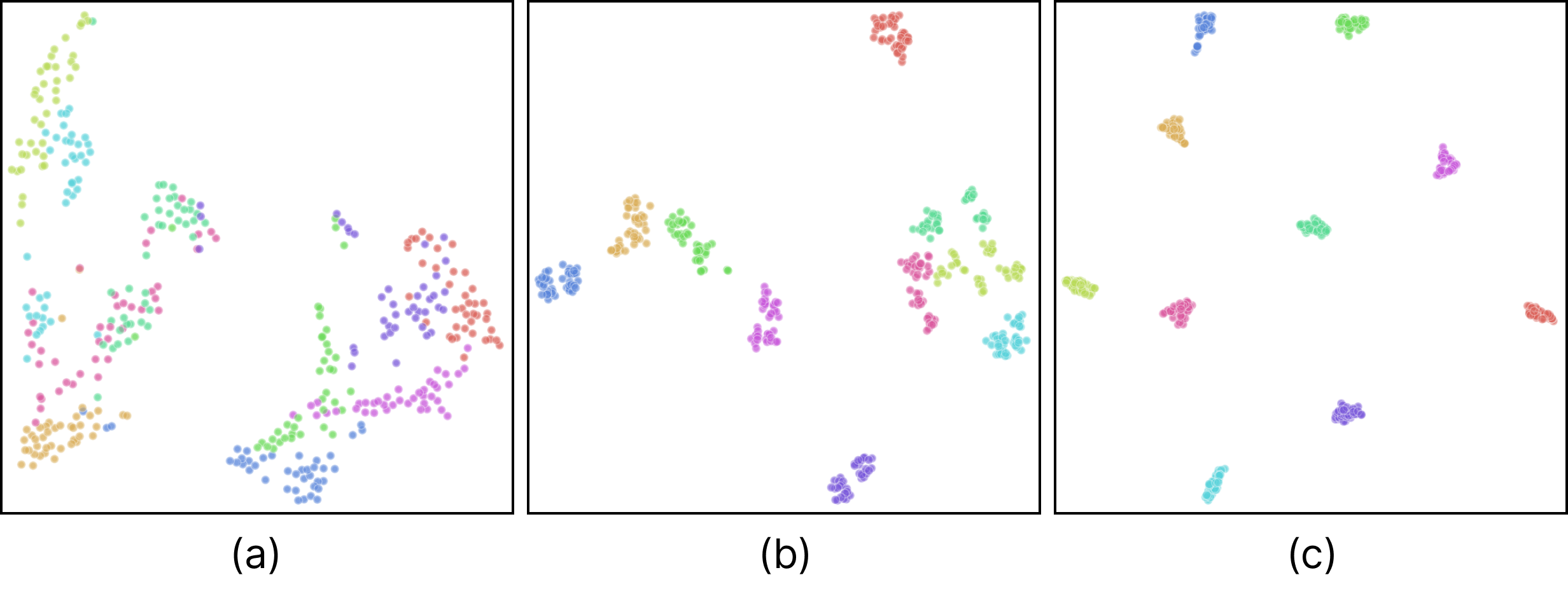}}
\vspace{-2mm}
\end{minipage}
\caption{t-SNE scatter plots of the recording environment embedding $z_r$. The different colors of the points represent different recording environments. (a) The acoustic environment embedding from \cite{acousticmatching}. (b) The baseline encoder in DiffRENT (c) The recording environment encoder in DiffRENT}
\label{fig:tsne}
\vspace{-2mm}
\end{figure}

\subsection{Subjective Evaluation}
Table \ref{table:subjective} shows the result of the subjective evaluation. While no significant differences were observed in objective evaluation, our model notably surpasses VoiceFixer in subjective evaluation. This result implies that our model generates sound with greater perceptual quality in comparison to VoiceFixer. Across all metrics, our model exhibits superior performance to the acoustic matching network. We found that performance of the acoustic matching network deteriorates when there is a substantial difference between the environments of the target and the content speech, while our model maintains consistent performance.

\subsection{Analysis of the Recording Environment Encoder}
We investigated the recording environment encoder to understand the behavior better by visualizing latent spaces with t-SNE. We used 400 audio files composed of 4 unseen speakers, 10 excerpts, and 10 environments. Test environments include a clean environment, environments with one unseen condition, and a totally unseen environment. Fig. \ref{fig:tsne} shows the t-SNE plots of each embedding from the acoustic embedding network \cite{acousticmatching} and two recording environment encoders of DiffRENT. Points sharing the same color represent audio within the identical recording environment. The point locations show that two encoders in DiffRENT disentangle the recording environment better than the acoustic embedding network. Furthermore, the encoder adapting ECAPA-TDNN exhibits more vivid isolation among different recording environments than the baseline encoder. 

\section{Conclusions}
\label{sec:conclusion}
We propose a novel recording environment transfer diffusion model for speech, which accurately models modeling microphone type and placement, room acoustics, and noise from a reference speech. The model effectively disentangles the recording environment and is adaptable to multiple scenarios of acoustic transform by the reference speech. Both objective and subjective evaluations show its effectiveness in speech enhancement and acoustic matching. Future work will involve enhancing the audio quality further by improving the neural vocoder. 
Additionally, we plan to develop a technique to reduce the number of refinement steps in the diffusion decoder for fast inference.




\vfill\pagebreak

\bibliographystyle{abbrv}
\bibliography{refs.bib}

\end{document}